\documentclass[namedreferences]{solarphysics}

\usepackage[hyperref,optionalrh,showbiblabels]{spr-sola-addons} % For Solar Physics 
\usepackage{graphicx}        % For eps figures, newer & more powerfull
\usepackage{color}           % For color text: \color command
\usepackage{breakurl}        % For breaking URLs easily trough lines
\newcommand{\aap}{    {\it Astron. Astrophys.}}

\newcommand{\apj}{    {\it Astrophys. J.}}

\newcommand{\apss}{   {\it Astrophys. Space Sci.}} 
\newcommand{\cjaa}{   {\it Chin. J. Astron. Astrophys.}}

\newcommand{\pasj}{   {\it Pub. Astron. Soc. Japan}}

\newcommand{\solphys}{{\it Solar Phys.}}

\chardef\us=`\_

\begin{document}

\begin{article}
\begin{opening}

\title{Photospheric Magnetic Free Energy Density of Solar Active Regions\\ {\it Solar Physics}}

\author[addressref={aff1},corref,email={e-mail.hzhang@bao.ac.cn}]{\inits{Hongqi}\fnm{Hongqi}~\lnm{Zhang}}%\sep
%\author[addressref=aff1,email={e-mail.b@mail.com}]{\inits{F.}\fnm{Fisrt}~\lnm{Author-b}}%\sep
%\author[corref,email={e-mail.c@mail.com}]{\inits{S.}\fnm{Second}~\lnm{Author-c}}%\sep
%\author{\inits{T.}\fnm{Third}~\lnm{Author-x}}
%\author{\inits{}\fnm{}~\lnm{}\orcid{}}
%\author{P.~\surname{Author-a}$^{1}$\sep
%        E.~\surname{Author-b}$^{1}$\sep
%        M.~\surname{Author-c}$^{2}$      
%       }

%   \institute{$^{1}$ First affiliation
%                     email: \url{e.mail-a} email: \url{e.mail-b}\\ 
%              $^{2}$ Second affiliation
%                     email: \url{e.mail-c} \\
%             }
\address[id=aff1]{Key Laboratory of Solar Activity, National Astronomical Observatories, Chinese Academy of Sciences, Beijing 100012, China}
%\address[id=aff2]{Second affiliation}
%\address[id=aff3]{Third affiliation}

\runningauthor{H. Zhang}
\runningtitle{Magnetic energy in photosphere}

%\begin{document}

%\title{Photospheric Magnetic Free Energy Density of Solar Active Regions}

%\author{Hongqi Zhang}
%\affil{Key Laboratory of Solar Activity, National Astronomical Observatories, Chinese
%Academy of Sciences, Beijing 100012, China, \\ {E-mail: hzhang@bao.ac.cn}}

\begin{abstract}
We present the photospheric energy density of  magnetic fields in { two} solar active regions (one of them recurrent) inferred from observational vector magnetograms, and compare it with other available differently defined  energy parameters of magnetic fields in the photosphere. 
{ We analyze the magnetic fields in Active Regions NOAA 6580-6619-6659 and 11158.}
The { quantity} $\frac{1}{4\pi}{\bf B}_{n}\cdot{\bf B}_{p}$ is an important energy parameter that reflects the contribution of magnetic shear to the difference between the potential  (${\bf B}_{p}$) and the non-potential magnetic field (${\bf B}_{n}$), and also the contribution  to the free magnetic energy near the magnetic neutral lines in the active regions. It is found that the photospheric mean magnetic energy density shows clear changes  before the powerful solar flares in Active Region NOAA 11158,  which is consistently with the change of magnetic fields in the flaring lower atmosphere.
\end{abstract}


\end{opening}

\section{Introduction}

{ The relationship of non-potential magnetic fields in active regions to solar flares and coronal mass ejections (CMEs) is an interesting topic of study.}  
It is believed that the non-potential magnetic fields are generated inside the Sun and then transported from the sub-atmosphere into  interplanetary space. The  photospheric non-potential magnetic field formed in active regions is also important for triggering the powerful solar flares and CMEs \citep[\textit{cf.}][]{Schmieder94} because it is difficult to accurately estimate the topology of the magnetic lines of force in the high solar atmosphere without  any information  about the photospheric magnetic fields.  Moreover, the associated change in photospheric vector magnetic fields with powerful flares in active regions has been detected from  observations made with magnetographs \citep[cf.][]{Chen89}. This means that  the analysis of photospheric magnetic fields is very important for understanding  solar eruptions.

%The formation and development of non-potential magnetic fields in solar active regions have been noticed due to the relationship with solar eruptive phenomena. 
The magnetic shear inferred from observational photospheric vector magnetograms is an important parameter to provide information about the  magnetic non-potentiality in solar active regions \citep[cf.][]{Krall82, Lv93}, while the vertical electric current and current helicity inferred from vector magnetograms mainly provide information about the spatial non-uniformity and chirality of  magnetic fields in the photosphere \citep[cf.][]{Severny71,Hagyard81, Seehafer94,Zhang95}. 
The magnetic fields and their free energy above the photosphere can be inferred by photospheric vector magnetograms in the approximation of force-free magnetic fields \citep{Chandrasekhar61,Low82}. Many questions remain, however, about the analysis of the real  configuration of magnetic fields in the solar atmosphere because it is difficult to use spectral information to accurately measure the vector magnetic field above the photosphere. %It is believed that our knowledge of the solar magnetic fields still has been constructed in an limited range. 
In this situation, obtaining more information from the observational photospheric vector magnetic fields and corresponding parameters becomes very important.

In this articale, we  analyze the energy distribution of  magnetic fields in the photosphere inferred from the observational vector magnetograms in flare-producing active regions.

\section{Definition of Free Magnetic Energy Density}
\label{S-fed}
The energy released by solar flares and other explosive events relies on the accumulation of the free magnetic energy (non-potential magnetic energy), which is defined as the difference between the total magnetic energy ($E_o$) and the potential magnetic energy ($E_{p}$):
\begin{eqnarray}
\label{eq:MGshenergy}
\Delta E = E_o-E_{p}.   %\nonumber
\end{eqnarray}
{  This means that the free energy is defined with respect to the energy of the potential field %is that the unique potential field 
%which matches % a given distribution of vertical magnetic field has 
%the lowest possible energy for that distribution  
 that matches the distribution of the observed field's normal component.  This potential field has the lowest possible magnetic energy that is still consistent with the boundary condition %(for example see Priest, 2014).
\citep[for example see][]{Priest14}.}

The total magnetic energy of a field $\bf B$ is given in the form 
\begin{equation}
\label{eq:int}
E= \int  B^2 dV,    %\nonumber
\end{equation} 
which means that the magnetic energy is a three-dimensional integral quantity, such as in the solar atmosphere. 

\begin{figure}%[htbp]
\centerline{\includegraphics[width=0.7\textwidth,clip=]{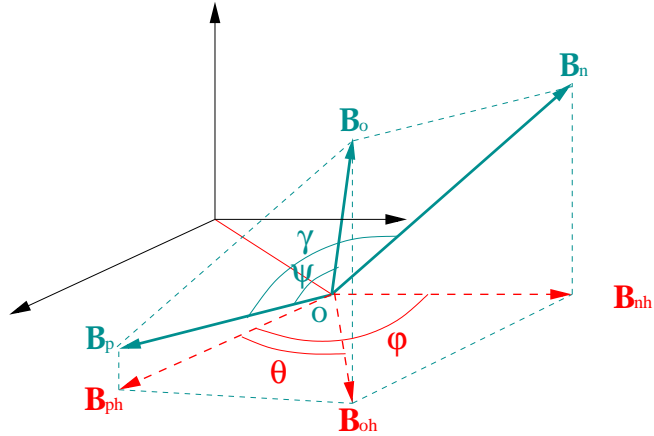}}
\caption{A schematic of the relationships among the  observed magnetic field ${\bf B}_o$, the potential field ${\bf B}_p$, and the non-potential field ${\bf B}_n$. The arrows in red indicate their respective horizontal components.  
}
\label{fig:MGsh0}
\end{figure}

\cite{Hagyard81} defined the source field to describe the non-potentiality of magnetic field on the photosphere:
\begin{eqnarray}
\label{eq:MGshear}
\emph{\textbf{B}}_{n}=\emph{\textbf{B}}_{o}-\emph{\textbf{B}}_{p},   %\nonumber
\end{eqnarray}
where $\emph{\textbf{B}}_{o}$ is the observed vector magnetic field, $\emph{\textbf{B}}_{p}$ denotes the potential field extrapolated from the { vertical }component of $\emph{\textbf{B}}_{o}$, and $\emph{\textbf{B}}_{n}$ is the so-called source field, that is, the non-potential component of the magnetic field.
By means of Equation (\ref{eq:MGshear}), it is found
\begin{equation}
\label{ }
B_n=\sqrt[]{B_o^2+B_p^2-2B_oB_p\cos\psi},
\end{equation}
where $\psi$ is the inclined angle between ${\bf B}_o$ and ${\bf B}_p$ in Figure \ref{fig:MGsh0}.

{ %Then state clearly that this paper basically only deals with 2D arrays. Later in the manuscript (p.3), the text refers to 
Now we can introduce the magnetic energy density $\rho$, 
%which might be introduced here, 
as in $\displaystyle E = \int \frac{B^2}{8\pi} dV  = \int \rho dV$, where the field energy density $\rho \equiv B^2/8 \pi$. Analogously, for $E_o$ and $E_p$, the observed and potential-field energy density are $\rho_o \equiv B_o^2/8 \pi$ and  $\rho_p \equiv B_p^2/8 \pi$, respectively.  We also note that the free energy density integrated over a volume $V$ yields an energy,  and the free energy density integrated over the photosphere yields a quantity with dimensions of energy per unit length.  This quantity is not, strictly speaking, a free energy, but it is plausibly related to the free energy present in the volume above the photosphere.  
The observation of photospheric vector magnetograms provides an opportunity to analyze the distribution of magnetic energy density in the low solar atmosphere. { We  exclusively} consider two-dimensional arrays of magnetic energy from photospheric magnetic fields in the following.} 

The energy density parameter of the non-potential magnetic field as defined by \cite{Lv93} is proportional to $\emph{\textbf{B}}_{n}^{2}$:
\begin{eqnarray}
\rho_{n}=\frac{(\emph{\textbf{B}}_{o}-\emph{\textbf{B}}_{p})^{2}}{8\pi}=\frac{\emph{\textbf{B}}_{n}^{2}}{8\pi}.    \label{Eq-fED}
\end{eqnarray}
%It also equals to the component form
%\begin{equation}
%\label{ }
%\rho_{free}^*=\frac{{B}_{nx}^{2}+{B}_{ny}^{2}+{B}_{nz}^{2}}{8\pi}.
%\end{equation}
This energy parameter had been also used to  analyze the solar  magnetic active cycles \citep[][]{Yangx12}.

It follows that the definition of the real free energy density (Equation (\ref{eq:MGshenergy})) is
\begin{eqnarray}
\rho_{free} = \rho_o - \rho_p=\frac{{\bf B}_{o}^2-{\bf B}_{p}^{2}}{8\pi}=\frac{({\bf B}_{n}+{\bf B}_{p})^{2}-{\bf B}_{p}^{2}}{8\pi},   
%\nonumber
\end{eqnarray}
and it can be written 
\begin{eqnarray}%\begin{equation}
\rho_{free}&=&\frac{1}{8\pi}{B}_n^2+\frac{1}{4\pi}{\bf B}_{n}\cdot{\bf B}_{p}=
\frac{1}{8\pi}{B}_n^2+\frac{1}{4\pi}{ B}_{n}{ B}_{p}\cos\gamma\nonumber\\ 
&=&\frac{1}{8\pi}{B}_n^2+\frac{1}{4\pi}B_{n}B_{p}\cos\left[\psi+\cos^{-1}\left(\frac{B_{o}-B_{p}\cos\psi}{B_{n}}\right)\right],
\label{eq:MGsh00}
\end{eqnarray}%\end{equation}
{ where $\gamma$ and $\psi$ are spatial angles, and $\cos^{-1}\left(\frac{B_{o}-B_{p}\cos\psi}{B_{n}}\right)=\gamma-\psi$, in Figure \ref{fig:MGsh0}.} 
We can find the relationship %in Figure \ref{fig:MGsh0} % \citep[see also][]{Lv93}  
\begin{eqnarray}
\label{ }
&&\cos\left[\psi+\cos^{-1}\left(\frac{B_{o}-B_{p}\cos\psi}{B_{n}}\right)\right]\nonumber\\
&&=\sin\vartheta_p\sin\vartheta_n
+\cos\vartheta_p\cos\vartheta_n\cos\left[\theta+\cos^{-1}\left(\frac{B_{oh}-B_{ph}\cos\theta}{B_{nh}}\right)\right] ,
\end{eqnarray}
{ where $\vartheta_n$ is the inclination angle between the vector of the non-potential field ${\bf B}_n$ and its horizontal component ${\bf B}_{nh}$, and $\vartheta_p$ is that between the vector of the potential field ${\bf B}_{p}$ and its horizontal component  ${\bf B}_{ph}$, while} the angle $\theta$ is defined as the { horizontal} angle between ${\bf B}_{oh}$ and ${\bf B}_{ph}$, and $\cos^{-1}\left(\frac{B_{oh}-B_{ph}\cos\theta}{B_{nh}}\right)=\varphi-\theta$, in Figure \ref{fig:MGsh0}.

Because we cannot { directly measure} the vertical component of the non-potential magnetic field in the photospheric layer from the observational vector magnetograms, 
{ we normally require a model field to match the observed vertical field, whether in the analysis of non-potential fields or potential fields.  For potential models, this
choice corresponds to the Neumann boundary condition \citep[see, for instance,][]{Sakurai82},   
{although }  
observations of changes in vertical magnetic fields associated with flares demonstrate that the observed vertical field is not generally consistent with the lowest energy state of the field.} 
This means that the vertical component of the magnetic field does not contribute to the magnetic free energy density from the photospheric vector magnetograms. { For observations near disk center, we can approximate the line-of-sight field as the vertical component of the observed field \citep{Sakurai82}.}
We choose to neglect the vertical component of the non-potential magnetic field in the following analysis.

As the { three-dimensional } shear of the magnetic field has been ignored \citep{Leka03}, the magnetic free energy density contributed by the horizontal components of these magnetic fields is
\begin{eqnarray}%\begin{equation}
\label{eq:MGsh0}
\rho_{fh}&=&\frac{1}{8\pi}({B}_{oh}^2-{B}_{ph}^2)\nonumber\\  
&=&\frac{1}{8\pi}({B}_{nh}^2+2{\bf B}_{nh}\cdot{\bf B}_{ph} )\\
&=&\frac{1}{8\pi}{B}_{nh}^2+\frac{1}{4\pi}B_{nh}B_{ph}\cos\varphi \nonumber\\
&=&\frac{1}{8\pi}{B}_{nh}^2+\frac{1}{4\pi}B_{nh}B_{ph}\cos\left[\theta+\cos^{-1}\left(\frac{B_{oh}-B_{ph}\cos\theta}{B_{nh}}\right)\right]\nonumber, 
\end{eqnarray}%\end{equation}
where $B_{oh}$, $B_{nh}$ and  $B_{ph}$ are the horizontal components of observational, non-potential, and potential magnetic field, respectively, and { horizontal } angles $\theta$ and $\varphi$ are defined in Figure \ref{fig:MGsh0}.  
Equation (\ref{eq:MGsh0}) implies that $\rho_{fh}$ is not necessarily positive {because the second term in  Equation (\ref{eq:MGsh0}) can be negative and its absolute value may be higher than the first one in some cases. 
From Equations (\ref{eq:MGsh00}) and (\ref{eq:MGsh0}), we also find the inclination angles between { the } observed vector magnetic field and { the} potential field do not simply { reflect the status of free energy density in the low solar atmosphere, regardless of whether we consider the contribution of the vertical component of the vector magnetic field.}
%  whether insight of the spatial or horizontal magnetic shear.  
}

The difference between the  free energy density $\rho_{fh}$ and the non-potential parameter  $\rho_{nh}$ contributed by horizontal components of magnetic fields is  
\begin{equation}
\label{eq:enegcomp1}
\bigtriangleup\rho_{eh}=\rho_{fh}-\rho_{nh}=\frac{1}{4\pi}{\bf B}_{nh}\cdot{\bf B}_{ph}.
%= \frac{1}{4\pi}(B_{nx}B_{px}+B_{ny}B_{py}).  
\nonumber
\end{equation}
%It  implies that the actual relationship between the magnetic field free energy and the shear angle of the field in the solar surface is relatively complex. 
This means that the contribution of the parallel component of the non-potential field relative to the potential field is non-negligible, and the  difference $\bigtriangleup\rho_{eh}$ between two differently defined magnetic energies will only vanish where the potential components of magnetic field are perpendicular to the non-potential field components. ${\bf B}_{oh}\cdot{\bf B}_{ph}$ relates to the normally defined magnetic shear and also to the potential field, where  ${\bf B}_{oh}\cdot{\bf B}_{ph}={\bf B}_{nh}\cdot{\bf B}_{ph}+B_{ph}^2$.  This means that ${\bf B}_{oh}\cdot{\bf B}_{ph}$  contains the contribution of the magnetic potential energy ($B_{ph}^2$) of solar active regions. 
We can find the shear angle $\theta$
\begin{equation}
\label{eq:shearangle}
\theta=cos^{-1}\left(\frac{B_{nh}B_{ph}cos\varphi+B_{ph}^2}{B_{oh}B_{ph}}\right)=cos^{-1}\left(\frac{B_{nh}cos\varphi+B_{ph}}{B_{oh}}\right).
\end{equation}
Based on this discussion, we can find that the normally defined shear angle $\theta$ cannot be used to conclusively determine information about the free magnetic energy in the lower solar atmosphere because Equation (\ref{eq:shearangle}) also partly relates to the term $B_{ph}^2$. { This is consistent with the idea that the magnetic shear contains information about the energy density that is contributed by the horizontal component of the potential field.
}

We know that the transverse components of the potential magnetic field are not an observed quantity, but  { have to} be inferred by extrapolating the longitudinal components of the observational magnetic field \citep[\textit{cf.}][]{Hagyard78, Sakurai82}. This is based on the assumption that magnetic lines of force extend from photospheric magnetic charges. This means that we cannot accurately observationally estimate the contribution of the longitudinal component of non-potential magnetic fields in solar active regions.

In the following, we analyze the different energy parameters based on the observed vector magnetic fields and the inferred potential magnetic fields in solar active regions. Based on the considerations above, we neglect the contribution of the longitudinal component of the magnetic field  when we study the magnetic energy in the lower solar atmosphere.

\section{Photospheric Magnetic Energy Parameters Contributed by Horizontal Components of Magnetic Fields in Active Regions}

To analyze the contribution of the magnetic energy density in the lower solar atmosphere by the horizontal components of magnetic field,
we introduce  Active Region NOAA 6580-6619-6659, which was observed in April-June 1991 at Huaioru Solar Observing Station, and NOAA 11158, which was observed  in February 2011  by the \textit{Heliospheric Magnetic Imager} (HMI) onboard the \textit{Solar Dynamics Observatory} (SDO). 
In this study, the transverse components of the potential magnetic field are calculated by extrapolating observational longitudinal components of magnetic fields in the active regions based on the  {  approximations that the line-of-sight flux is equivalent to the vertical flux, and that this flux is due to magnetic charge. }
  
\subsection{Recurrent Active Region NOAA 6580-6619-6659}

\begin{figure}%[htbp]
\begin{center}
\includegraphics[width=1\textwidth,clip=]{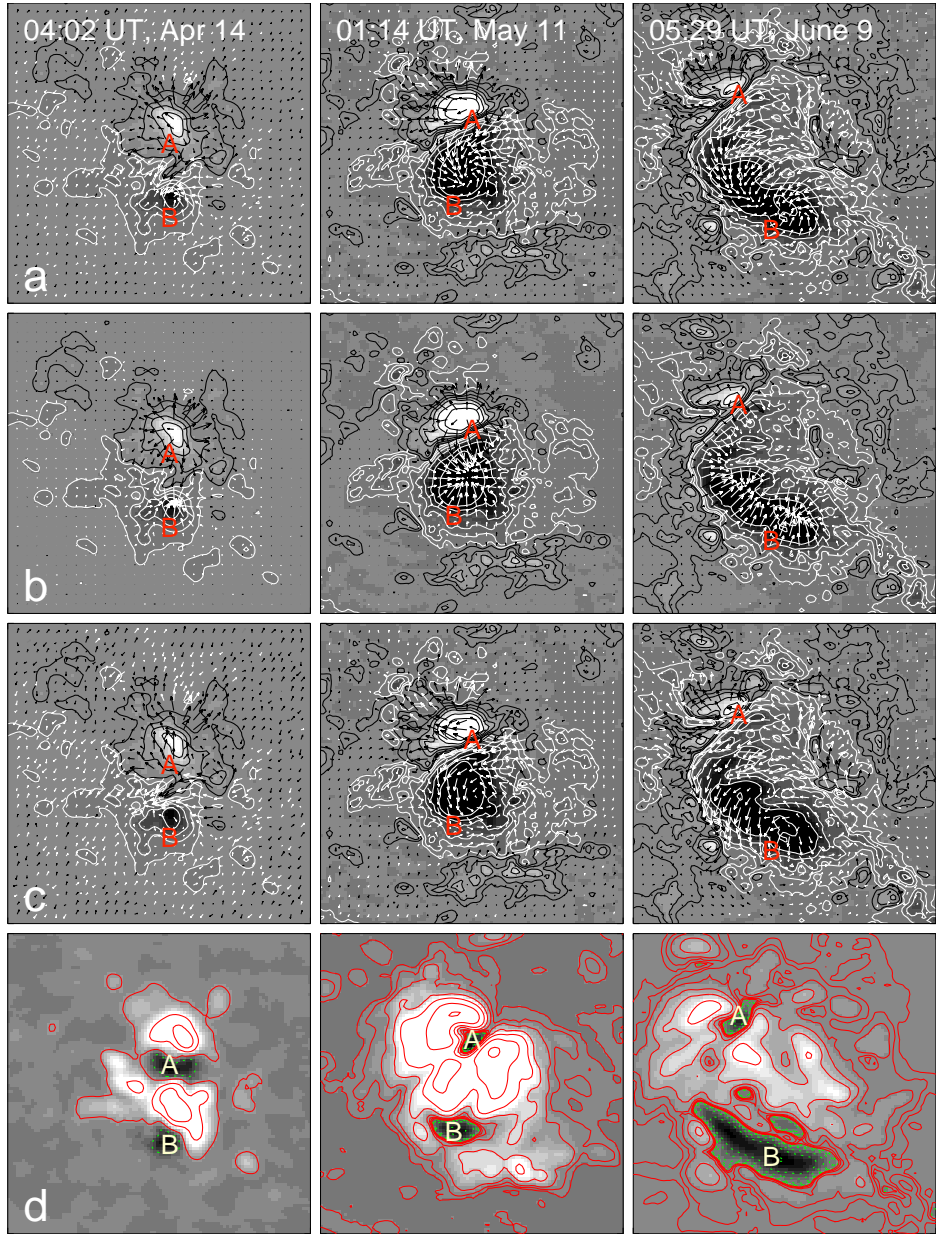}
\end{center}
\caption{The vector magnetic fields in Active Regions NOAA 6580 (N28, W12) on  April 14, 1991  (left), NOAA 6619 (N29, E09) on May 11, 1991 (middle) and NOAA 6659 (N32, E05) on June 9, 1991 (right).   Top row (a): The arrows mark the observed transverse component of field ${\bf B}_o$. Second row (b): The arrows show the transverse components of the potential  field ${\bf B}_p$ inferred from the longitudinal components of observational magnetic fields. Third row (c): The arrows show the transverse components of the non-potential field ${\bf B}_n$ inferred from Equation (\ref{eq:MGshear}).  The contours  indicate the longitudinal magnetic field distribution of $\pm$50, 200, 500, 1000, 1800, and 3000 G.   Bottom row (d):  The contours  show the distribution of  free magnetic energy densities $\rho_{fh}$ of $\pm$1, 5, 10, 20, 50, 100, 200, and 400$\times 10^3 $ (Mx$^2$cm$^{-4})$ 
{for the quantity shown in  gray scale}, where  solid (dashed) lines correspond to positive (negative) sign.}
\label{Fig:nonpoten}
\end{figure}

Active Region NOAA 6580 occurred on the solar disk in April 1991, NOAA 6619 in  May 1991 and NOAA 6659 in  June1991. This was a super delta active region on the solar surface  in a sequence of solar rotations.  A series of photospheric vector magnetograms of this active region was observed by the video vector magnetograph at Huairou Solar Observing Station of National Astronomical Observatories, Chinese Academy of Sciences \citep{Ai86}. 
Row (a) of Figure \ref{Fig:nonpoten} shows observed vector magnetograms of NOAA 6580  on 14 April 1991 (N28W10), NOAA 6619 on  11 May 1991 (N29E11) and NOAA 6659 on 9 June 1991 (N32E06).  Therefore this active region can be called NOAA 6580-6619-6659.  
A series of powerful flares erupted  in this  super active region  in Solar Cycle 22 \citep[cf.][]{Sakurai92, Bumba93, Liu96, Ji98, Ji03, Schmieder94, Zhang94a, Zhang94b, Yan95a, Yan95b, Zhang95, Zhang96, Zhang01, Wang97, Wang06}.

\begin{figure}%[htbp]
\begin{center}
\includegraphics[width=1\textwidth,clip=]{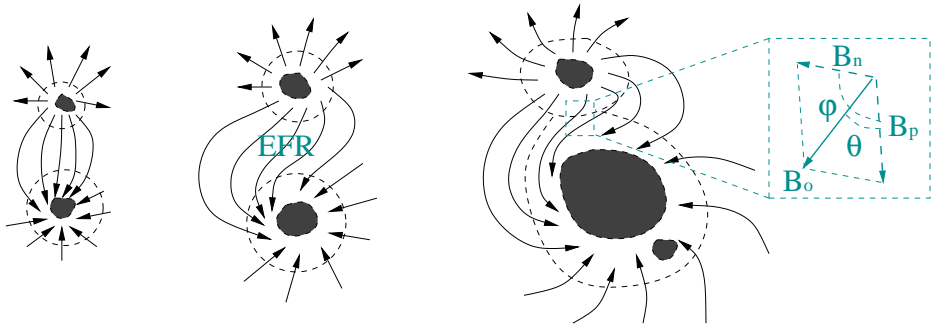}
\end{center}
\caption{A schematic of the development of Active Region NOAA 6580-6619-6659 in June 1991, and also the relationship among the  observed magnetic field ${\bf B}_o$, the potential field ${\bf B}_p$, and the non-potential field ${\bf B}_n$  as viewed from above.   }
\label{fig:MGsh1}
\end{figure}

\begin{figure*}[htbp]
\begin{center}
\includegraphics[width=1\textwidth,clip=2]{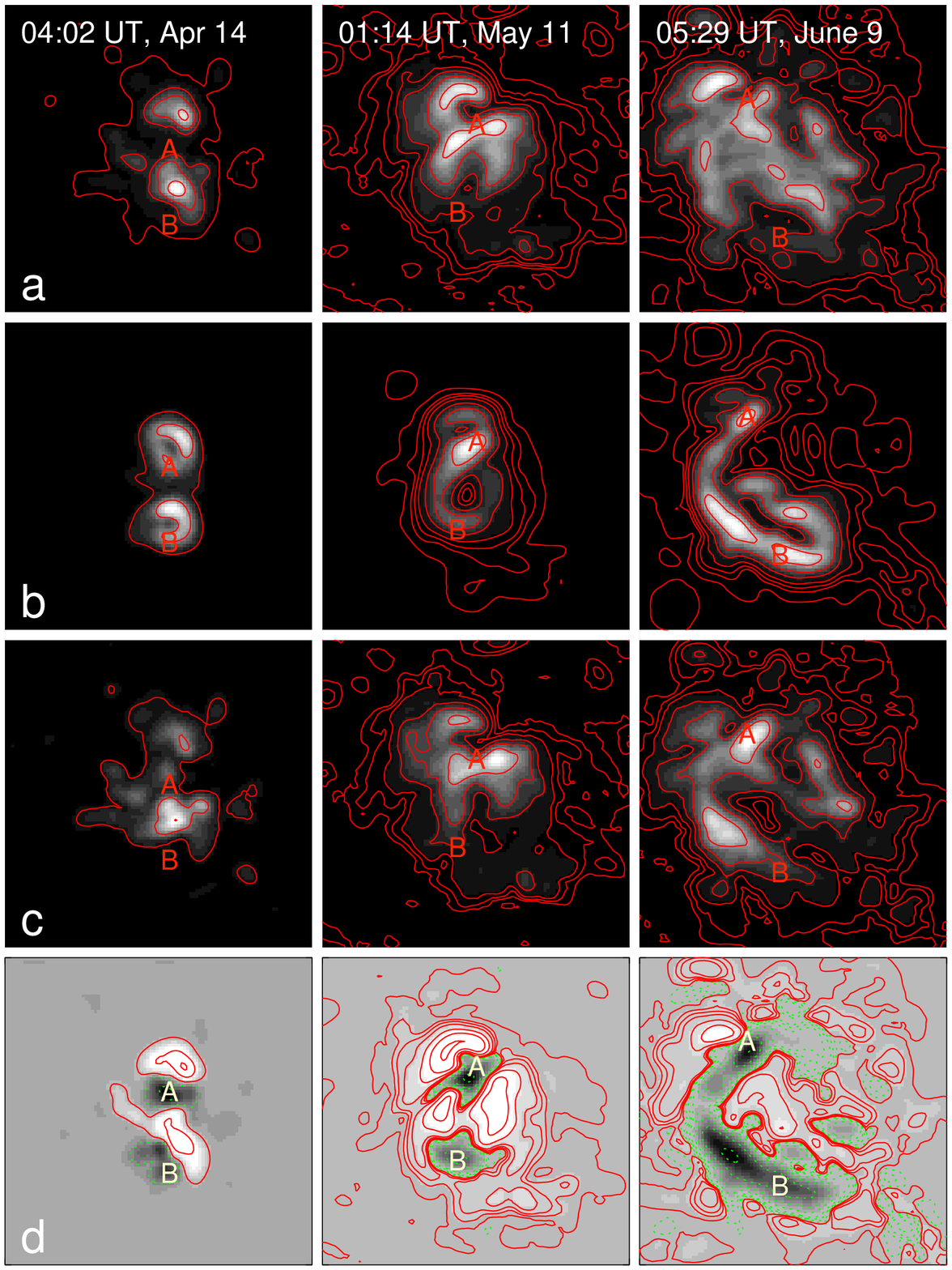}
\caption{The magnetic energy density parameters in Active Region NOAA 6580 on April 14, 1991 (left), NOAA 6619 on May 11, 1991 (middle), and NOAA 6659 on June 9, 1991 (right).   Top row (a): Inferred from the observed transverse component of field ${\bf B}_o$. Second row (b):  Inferred from the  the transverse components of  potential field ${\bf B}_p$ calculated by the longitudinal components of observed magnetic fields. Third row (c):  Inferred from the transverse components of  non-potential field ${\bf B}_n$.  Bottom row (d):  The difference $\frac{1}{4\pi}{\bf B}_{nh}\cdot{\bf B}_{ph}$ between magnetic free energy density  $\rho_{fh}$  and  the energy density of  the non-potential field $\rho_{nh}$  in Equetion (\ref{eq:enegcomp1}).  The  contours indicate $\pm$1, 5, 10, 20, 50,100, 200, and 400$\times 10^3$ (Mx$^{2}$ cm$^{-4}$)
{ for the quantity shown in  gray scale}, where  solid (dashed) lines correspond to positive (negative) sign.}
\label{Fig:nonpotene}
\end{center}
\end{figure*}

Figure \ref{Fig:nonpoten} also shows  the relationship among the observed highly twisted vector magnetic fields, the calculated potential and non-potential magnetic fields,   and the corresponding magnetic free energy densities in the super active region NOAA 6580-6619-6659  in 1991.  We can find in row (a) of Figure \ref{Fig:nonpoten}  that  highly twisted transverse magnetic fields form in the active region. Row (b) of Figure \ref{Fig:nonpoten} shows the calculated transverse components of potential magnetic fields inferred by the observed longitudinal magnetic fields of the active region. Row (c) of Figure \ref{Fig:nonpoten} shows the non-potential components of magnetic  fields inferred by Equation (\ref{eq:MGshear}). In row (d) of Figure \ref{Fig:nonpoten}, the magnetic free energy densities in the active region are inferred by  $\rho_{fh}$   in Equation (\ref{eq:MGsh0})  to be contributed by the transverse components of fields alone. We find that the free energy density is  negative in some areas (highly sheared magnetic neutral lines) of the active region, i.e. $(\frac{1}{8\pi}{B}_{nh}^2+\frac{1}{4\pi}{\bf B}_{nh}\cdot{\bf B}_{ph})<0$ in Equation
(\ref{eq:MGsh00}), where  ${\bf B}_{ph}$ is inferred by the observed longitudinal magnetic field. These areas  of negative sign, labeled $A$ and $B$, can be defined as negative energy regions in row (d) of Figure \ref{Fig:nonpoten}. 

Figure \ref{fig:MGsh1} shows a schematic  of the development process of the observed highly twisted vector magnetic field in the super active region NOAA 6580-6619-6659  in 1991.  With the emergence of the new magnetic flux, the highly sheared  magnetic field formed near the magnetic neutral line of the active region, and the transverse component of the magnetic fields gradually became parallel to the magnetic neutral line. This relates to  the storage of free magnetic energy and the formation of `negative energy regions' in the active region.

Figure \ref{fig:MGsh1} also shows the relationship among the  observed magnetic field ${\bf B}_o$, the potential field ${\bf B}_p$, and the non-potential field ${\bf B}_n$ near the magnetic neutral line in the active region {as viewed from above}. This provides a possibility for the occurrence of negative energy regions in the active regions as  $\frac{1}{8\pi}{B}_n^2<-\frac{1}{4\pi}{\bf B}_{n}\cdot{\bf B}_{p}$.  The relevant results can be found in Figure  \ref{Fig:nonpoten}d. We can find that in these  negative energy regions the inclined angles $\varphi$ between the potential and non-potential components of transverse fields in the active region are larger than 90$^\circ$, i.e. $\frac{1}{4\pi}{\bf B}_{n}\cdot{\bf B}_{p}$ is  negative {in Figures \ref{Fig:nonpoten} and \ref{fig:MGsh1}}.

For comparison, Figure \ref{Fig:nonpotene} shows the distribution of different types of  magnetic energy density parameters in Active Region NOAA 6580-6619-6659  in 1991.  In the calculation of the photospheric energy density, only  transverse components of the magnetic fields were used because the longitudinal components do not change in this analysis. 
Row (a) of Figure \ref{Fig:nonpotene} shows the magnetic energy densities inferred from the observational transverse magnetic fields.  
Row (b) of Figure \ref{Fig:nonpotene} shows the magnetic energy density inferred from the potential transverse magnetic fields, which is calculated using only  the longitudinal components of magnetic fields. Row (c) of Figure \ref{Fig:nonpotene} shows the energy density parameters $\rho_{n}$ of non-potential  magnetic fields.
 
There are obvious differences  among the magnetic energy density parameters inferred from the  observational  transverse components of fields ${\bf B}_{oh}$,  from the potential transverse components of field ${\bf B}_{ph}$, and   from the non-potential transverse components of field ${\bf B}_{nh}$.  We note that row (d) of Figure \ref{Fig:nonpotene} shows the difference $\frac{1}{4\pi}{\bf B}_{nh}\cdot{\bf B}_{ph}$  between the magnetic energy density  $\rho_{fh}$ (Equation (\ref{eq:MGsh0})) and the energy density  parameter $\rho_{n}$ of the non-potential  transverse magnetic fields  in Equation (\ref{Eq-fED}).  The large-scale negative sign areas  $A$ and $B$  reflect where components of the potential field show large inclination angles to the non-potential field in Active Region NOAA 6580-6619-6659.

\begin{figure*}[htbp]
\begin{center}
\includegraphics[width=1.\textwidth,clip=2]{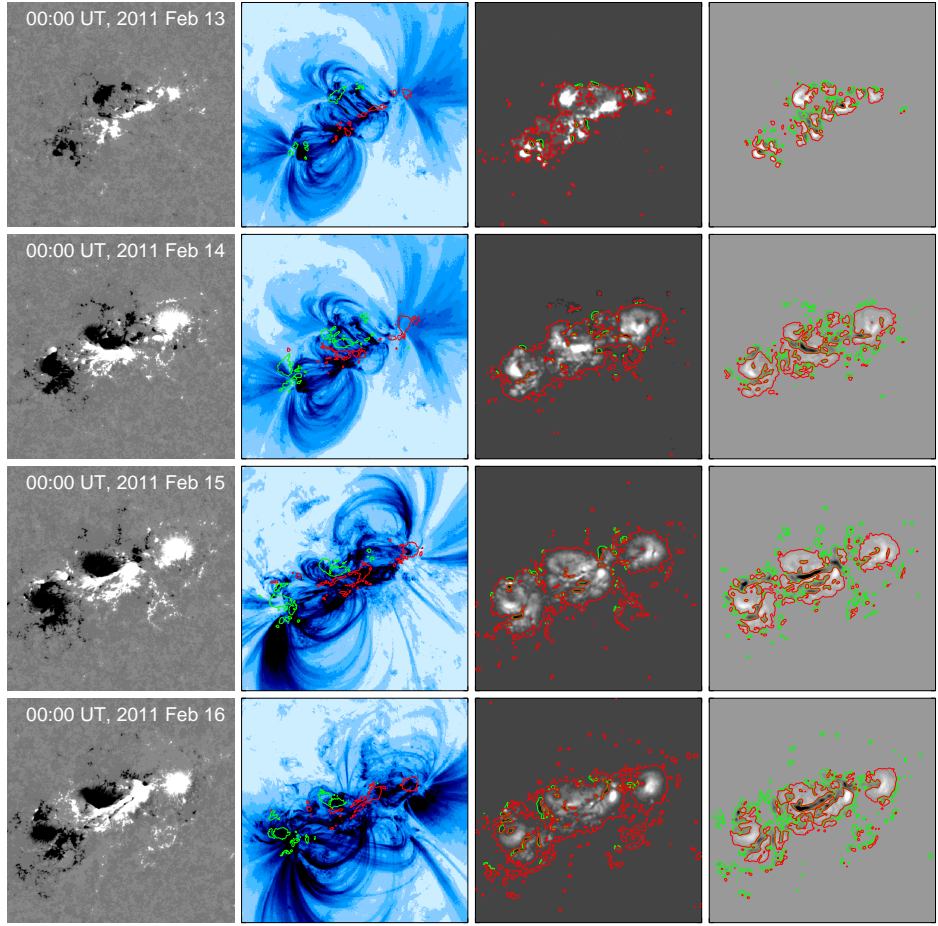}
\caption{
Active Region NOAA 11158 in  February 2011 at four different times. From left to right: the longitudinal magnetic fields, 171 \AA{} images overlaid with contours of $\pm10^3$ G magnetic fields, the magnetic free energy density $\rho_{fh}$, and  the quantity $\frac{1}{4\pi}{\bf B}_{nh}\cdot{\bf B}_{ph}$.  The contours of the magnetic energy densities indicate $\pm10^3$ (Mx$^{2}$ cm$^{-4}$). The  red (green) contours refer to the positive (negative) sign.
}
\label{Fig:magene11158}
\end{center}
\end{figure*}

\begin{figure*}[htbp]
\begin{center}
\vspace{3mm}
\includegraphics[width=0.6\textwidth,clip=2]{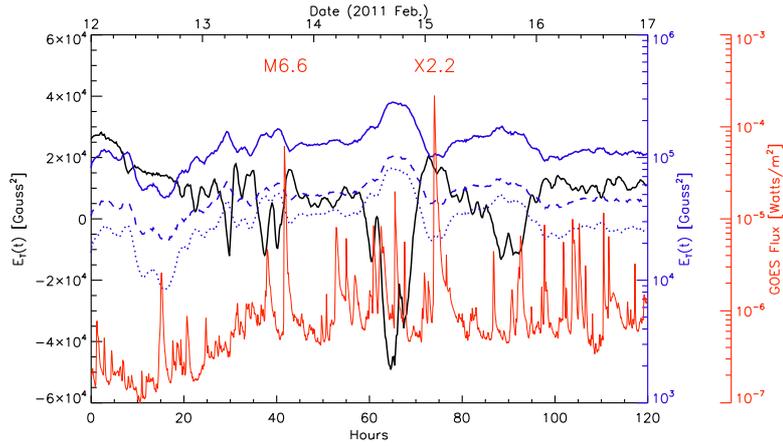}
\end{center}\caption{The evolution of mean magnetic energy density parameters in Active Region NOAA 11158 in February 2011.  The blue solid line ($\overline{B_{oh}^2/8\pi}$) is calculated based on the observational photospheric transverse field, the blue dotted line ($\overline{B_{ph}^2/8\pi}$) is the calculated transverse potential field, and the blue dashed line is the $\overline{\rho_{nh}}=\overline{B_{nh}^2/8\pi}$ inferred from the transverse non-potential field, while the black solid line shows the corresponding $\overline{\frac{1}{4\pi}{\bf B}_{nh}\cdot{\bf B}_{ph}}$.  The red line shows the GOES flux.  }
\label{Fig:enelines11158}
\end{figure*}

\begin{figure*}[htbp]
\begin{center}
\includegraphics[width=1.\textwidth,clip=2]{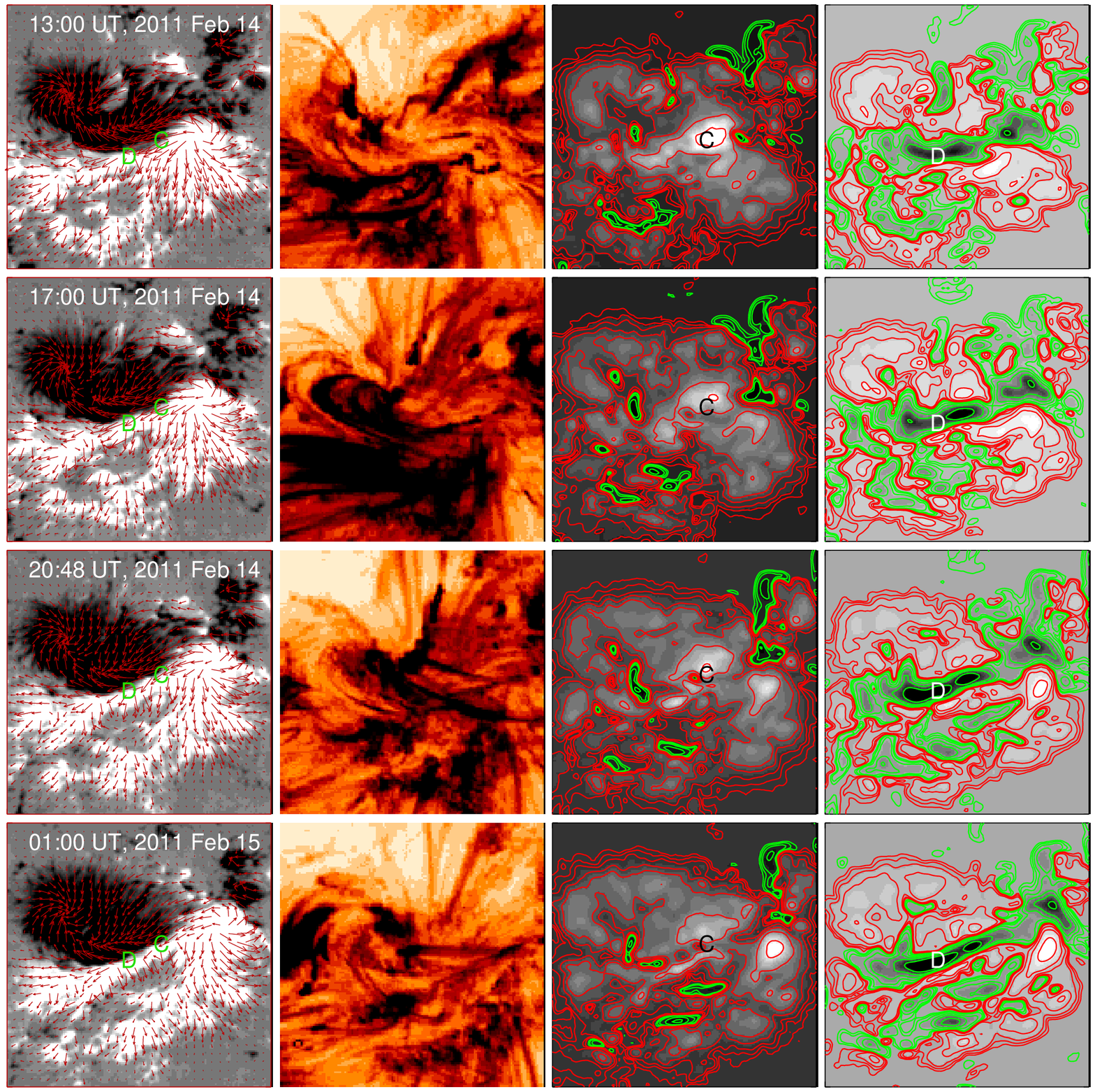}
\caption{
A local area of Active Region NOAA 11158 in February 14-15, 2011. From left to right:  longitudinal magnetic fields, 171 \AA{} images,  magnetic free energy density  $\rho_{fh}$, and  the quantity $\frac{1}{4\pi}{\bf B}_{nh}\cdot{\bf B}_{ph}$.  The contours of the magnetic energy density indicate  $\pm$1, 5, 10, 20, 50, 100, 200, and 400$\times 10^3  $ (Mx$^{2}$ cm$^{-4}$). The  red (green) contours refer to the positive (negative) sign.
}
\label{Fig:maen11158-14-15}
\end{center}
\end{figure*}

\begin{figure*}[htbp]
\begin{center}
\includegraphics[width=1.\textwidth,clip=2]{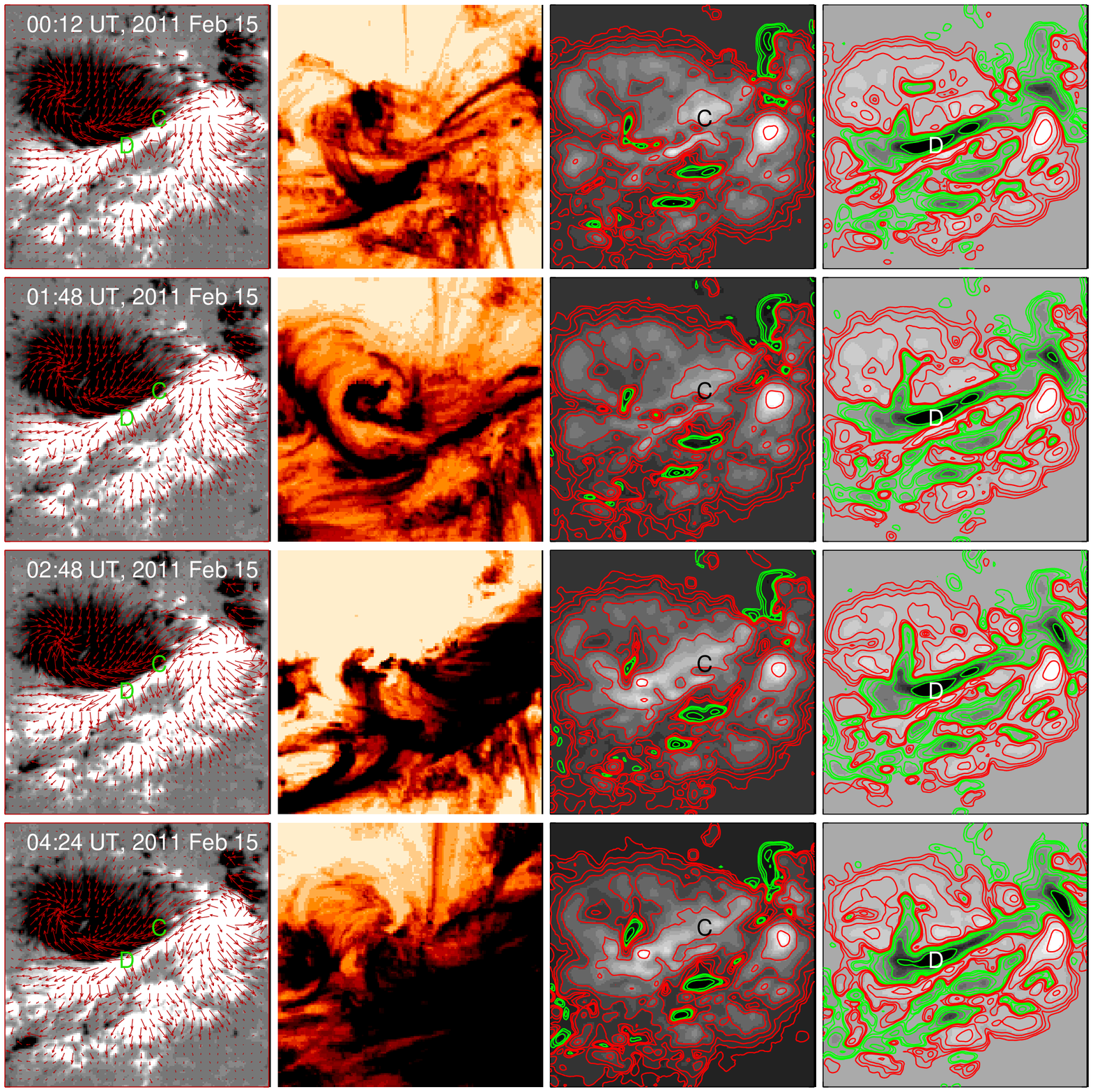}
\caption{
A local area of Active Region NOAA 11158 on February 15,  2011. From left to right:  longitudinal magnetic fields, 171\AA{} images,  magnetic free energy density  $\rho_{fh}$, and the difference  quantity $\frac{1}{4\pi}{\bf B}_{nh}\cdot{\bf B}_{ph}$.  The contours of the magnetic energy density indicate  $\pm$1, 5, 10, 20, 50, 100, 200, and 400$\times 10^3$  (Mx$^{2}$ cm$^{-4}$). The  red (green) contours refer to the positive (negative) sign.
}
\label{Fig:maen11158-15}
\end{center}
\end{figure*}

\subsection{Active Region NOAA 11158}

It is notable that vector magnetic field data have been made available with the \textit{Helioseismic and Magnetic Imager} (HMI)
instrument \citep{Schou12} onboard the  \textit{Solar Dynamics Observatory} (SDO). 
Active Region NOAA 11158  is a notable active region observed by HMI.
A series of articles has focused on the development of magnetic fields and solar flares and CMEs in Active Region NOAA 11158 based on observational vector magnetograms and  EUV images \citep[cf.][]{Aschwanden14, Bamba13,  Chintzoglou13, Dalmasse13, Gary14, Gosain12, Guerra15, Inoue13, Jiang12, Jing12,   LiuSc12, Maurya12, Malanushenko14, Kazachenko15, Petrie12, Song13, Su12,  Tarr13, Toriumi13,   Tziotziou13, Vemareddy12a, Vemareddy12b, Wang12, Wang13, Yang14, Zhao14, Zhang14}.  The development of photospheric vector magnetic fields in the active region with flares and CMEs is also related to the basic question of the distribution of magnetic energy in the photosphere and the storage process of magnetic energy for the powerful flares and CMEs. 

\cite{Fisher98} found that the X-ray luminosity shows the best correlation with the total unsigned magnetic flux.
Similarly, \cite{Fludra08} reported  EUV emission to be linearly related to the amount of magnetic flux, regardless of potentiality.  
Figure \ref{Fig:magene11158} shows  the distribution of  magnetic field, 171\AA{} images, free magnetic energy density $\rho_{fh}$ (Equation (\ref{eq:MGsh0})) and  the difference  $\frac{1}{4\pi}{\bf B}_{nh}\cdot{\bf B}_{ph}$  between magnetic energy density  $\rho_{fh}$  and energy density parameter   $\rho_{n}$  (Equation (\ref{Eq-fED})) of the non-potential  magnetic field. 
Similar to Active Region  NOAA 6580-6619-6659, we also find some small-scale negative sign areas of free magnetic energy density in Active Region NOAA 11158. 

Figure \ref{Fig:enelines11158} shows the evolution of the mean values of different energy parameters of Active Region NOAA 11158 with the GOES flux inferred from a series of vector magnetograms in the period of February 12-16, 2015.  It is found that the mean magnetic free energy density $\overline{\rho_{fh}}$ (Equation (\ref{eq:MGsh0})), the mean energy density parameter   $\overline{\rho_{nh}}$  (Equation (\ref{Eq-fED})) of the non-potential  magnetic field, and the mean potential energy density in Active Region NOAA 11158 in February 2011  show almost the same evolution tendency. The photospheric mean magnetic energy densities of the active region tend to decrease before the X-ray M6.6 flare on  February 13 and the X2.2 flare on February 15. Furthermore, the difference $\overline{\frac{1}{4\pi}{\bf B}_{nh}\cdot{\bf B}_{ph}}$ (between the mean magnetic free energy density  $\overline{\rho_{fh}}$  and the mean energy density parameter  $\overline{\rho_{nh}}$   of the non-potential  magnetic field)  shows the opposite trend relative to others. This suggests the possibility that some amount of free magnetic energy is stored in the lower solar atmosphere before  powerful flares.  This is also consistent with the release of  free magnetic energy driving  powerful flares.  
%We should also need to notice the similar evidence on the  evolution of other magnetic parameters with powerful flares, such as total unsigned magnetic flux and integrated unsigned vertical current density, had been analyzed  \citep[e.g.][]{Schrijver07,Sun12}.  
%Moreover,  observational errors are also notable, 
{ Moreover, the limitations on the time resolution of the magnetic field observations should also be noted,}
for example,  HMI vector magnetograms incorporate data integrated over 1215 seconds \textit{etc.} \citep{Liuetal12}, which implies that some changes do not provide sufficient evidence to show if the evolution of the magnetic field actually precedes the flares. 

Figure \ref{Fig:maen11158-14-15} shows the evolution of the vector magnetic field, 171 \AA{} images, magnetic free energy density $\rho_{fh}$ and the difference quantity $\frac{1}{4\pi}{\bf B}_{nh}\cdot{\bf B}_{ph}$  in the middle of Active Region NOAA 11158 to analyze the evolution of these quantities before the X2.2 flare on February 15.  The large-scale negative energy density parameter $\frac{1}{4\pi}{\bf B}_{nh}\cdot{\bf B}_{ph}$ tends to { extend along} the direction of EUV 171 \AA{} loops and the highly sheared  major magnetic neutral line between large-scale opposite polarities in the active region. As compared with the evolution of the magnetic energy density, it is found that the maximum value  of the free magnetic energy density $C$ weakens, and the  channel where $\frac{1}{4\pi}{\bf B}_{nh}\cdot{\bf B}_{ph}$ is negative (marked by  $D$) tends to increase near the magnetic neutral line in the Active Region NOAA 11158 before the X2.2 flare.

Figure \ref{Fig:maen11158-15} shows the change in the vector magnetic fields, 171  \AA{} images, magnetic free energy density $\rho_{fh}$, and the difference quantity $\frac{1}{4\pi}{\bf B}_{nh}\cdot{\bf B}_{ph}$  in the local area of Active Region NOAA 11158 { from 90 minutes before  the X2.2 flare (start at 01:44, peak at 01:45, end at 01:56 UT) to after } on February 15,  2011.  The eruption of EUV 171 \AA{} loops occurred near the highly sheared magnetic neutral line and also above the extreme negative $\frac{1}{4\pi}{\bf B}_{nh}\cdot{\bf B}_{ph}$ area.         

\begin{figure*}[htbp]
\begin{center}
\includegraphics[width=0.8\textwidth,clip=2]{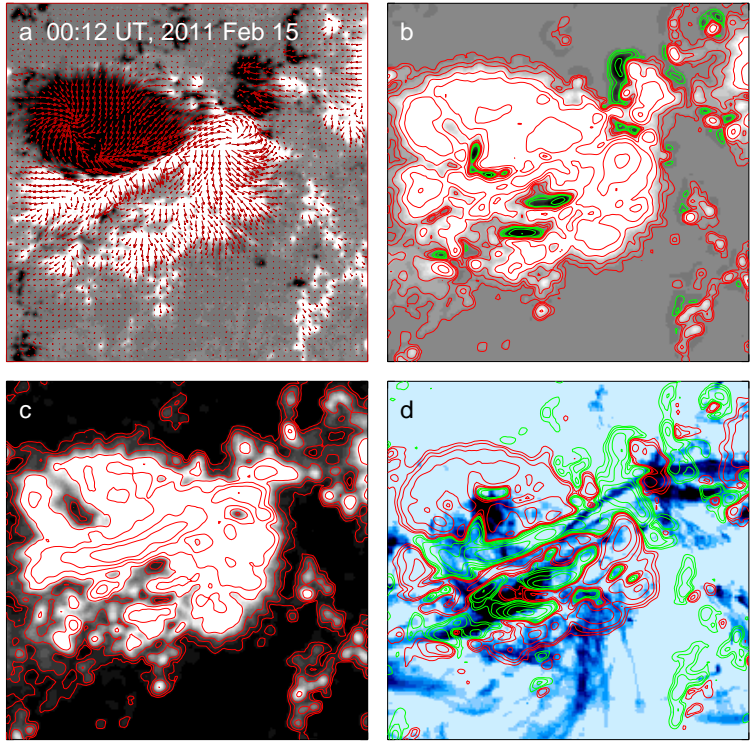}
\caption{(a) The vector magnetic field in Active Region NOAA 11158 on February 15,  2011. (b) The magnetic free energy density  $\rho_{fh}$ (Equation (\ref{eq:MGsh0})) and (c)  the energy density of the non-potential  magnetic field  $\rho_{n}$  (Equation (\ref{Eq-fED})) in Active Region NOAA 11158. (d) The difference $\frac{1}{4\pi}{\bf B}_{nh}\cdot{\bf B}_{ph}$ (between the magnetic free energy density  $\rho_{fh}$  and the energy density parameter $\rho_{nh}$   (Equation (\ref{eq:enegcomp1}))  of the non-potential  magnetic field)  overlaid with the AIA 171 \AA{} image. The contours indicate the magnetic energy density of $\pm$1, 5, 10, 20, 50, 100, 200, and 400$\times 10^3$ (Mx$^{2}$ cm$^{-4}$), where  red (green) contours refer to the positive (negative) sign.  }
\label{Fig:nonpotene11158e}
\end{center}
\end{figure*}

For a detailed analysis, Figure \ref{Fig:nonpotene11158e} shows the relationship between the vector magnetic field and different energy parameters in Active Region NOAA 11158 at 00:12 UT on  February 15, 2011. Figures \ref{Fig:nonpotene11158e}b and  \ref{Fig:nonpotene11158e}c show the magnetic free energy density  $\rho_{fh}$ (Equation (\ref{eq:MGsh0})) and the energy density parameter  $\rho_{nh}$  (Equation (\ref{Eq-fED}))  of non-potential  magnetic field in Active Region NOAA 11158 in February 2011, respectively.   Figure \ref{Fig:nonpotene11158e}d  shows $\frac{1}{4\pi}{\bf B}_{nh}\cdot{\bf B}_{ph}$ overlaid with 171 \AA{}  images. The large-scale extreme negative value of $\frac{1}{4\pi}{\bf B}_{nh}\cdot{\bf B}_{ph}$ tends to form along the direction of EUV 171 \AA{} loops and the highly sheared magnetic major neutral line between large-scale opposite polarities in the active region.
This is important for understanding the distribution of  magnetic energy above the solar surface.

\begin{figure*}[htbp]
\begin{center}
\includegraphics[width=0.8\textwidth,clip=2]{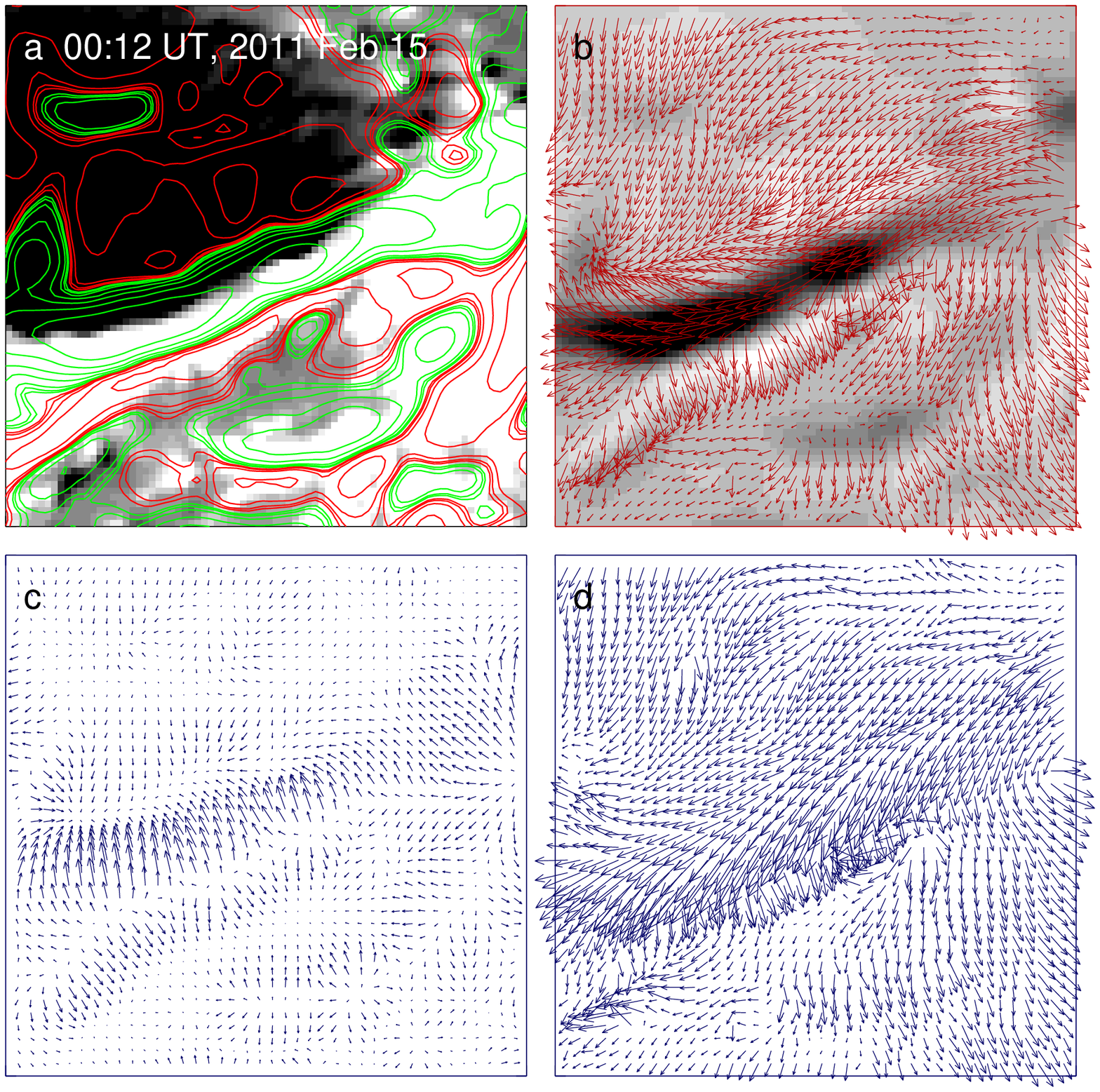}
\caption{
(a) A local area of longitudinal magnetic field (gray scale) and the  corresponding $\frac{1}{4\pi}{\bf B}_{n}\cdot{\bf B}_{p}$ (contours) in Active Region NOAA 11158 on February 15,  2011 near the center of Figure \ref{Fig:nonpotene11158e}. The contours indicate the magnetic energy density of $\pm$1, 5, 10, 20, 50, 100, 200, and 400$\times 10^3  $ (Mx$^2$cm$^{-4}$), where  red (green) contours refer to the positive (negative) sign. (b) The $\frac{1}{4\pi}{\bf B}_{n}\cdot{\bf B}_{p}$ (gray scale) and  the corresponding observed transverse magnetic field.  
(c) The corresponding potential component  and (d) the non-potential component of the transverse field.
}
\label{Fig:nonpotene11158f}
\end{center}
\end{figure*}

Figure \ref{Fig:nonpotene11158f} shows the relationship between the energy density parameter $\frac{1}{4\pi}{\bf B}_{nh}\cdot{\bf B}_{ph} $ and the different components of the vector magnetic field in the local area of Figure \ref{Fig:nonpotene11158e} in Active Region NOAA 11158. It also shows that negative-sign areas of $\frac{1}{4\pi}{\bf B}_{nh}\cdot{\bf B}_{ph}$ formed near the magnetic neutral lines with highly sheared transverse magnetic fields, where the most significant difference between the non-potential and potential fields can be found.

{ In addition to the analysis of other magnetic parameters, such as magnetic flux near neutral lines \citep[e.g.][]{Schrijver07} and integrated unsigned vertical current density \citep[e.g.][]{Sun12},
Figure \ref{Fig:maen11158-curhel} shows the distribution of the vertical current (${\bf J}_z$) and current helicity density ($(\nabla\times{\bf B})_z\cdot{\bf B}_z$) in the  middle of Active Region NOAA 11158 inferred from observed vector magnetograms with the comparison of the magnetic free energy density  $\rho_{fh}$ and  the difference  quantity $\frac{1}{4\pi}{\bf B}_{nh}\cdot{\bf B}_{ph}$.  These magnetic energy parameters are presented in Figure \ref{Fig:maen11158-14-15} with the relevant discussions. We can find that the magnetic energy parameters show obvious different density distributions relative to the current and current helicity densities because energy parameters reflect the storage of magnetic energy in the solar atmosphere, while the current and current helicity density provide information on the twist and handedness of magnetic fields in active regions. These show different aspects of magnetic lines of force in solar active regions, even though they are also important quantities in the analysis of solar eruptive activities.
Moreover, the development of the vector magnetic field and its relationship with helicities and  spectrums in the active region are discussed by \cite{Zhang14,Zhang16}.

\begin{figure*}[htbp]
\begin{center}
\includegraphics[width=0.8\textwidth,clip=2]{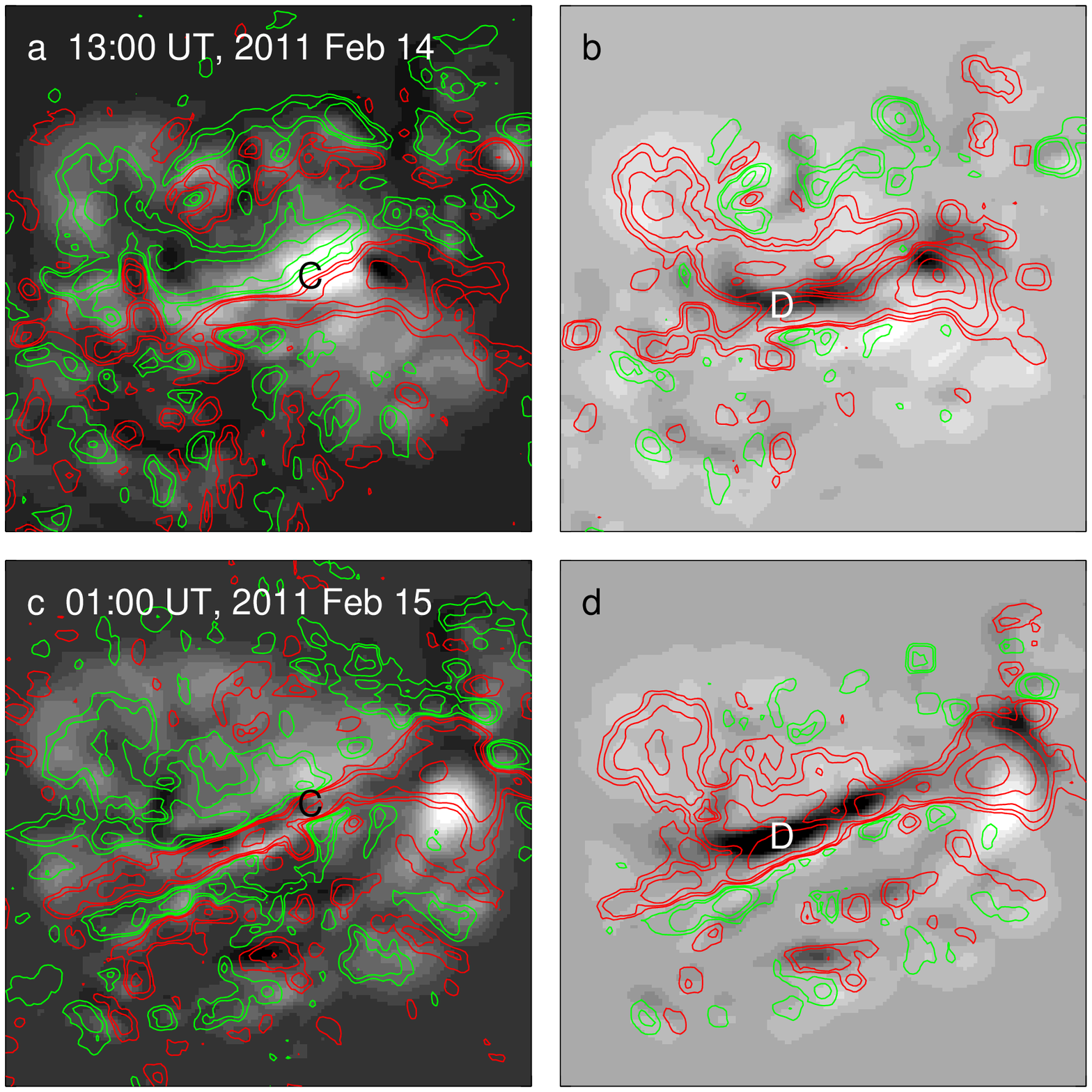}
\caption{
The vertical current (${\bf J}_z$) and current helicity density ($h_z=(\bigtriangledown\times{\bf B})_z\cdot{\bf B}_z$) in Active Region NOAA 11158 on February 14,  2011 (top) and February 15, 2011 (bottom). 
The contours in the left panels (a and c) indicate the vertical current density of $\pm$5, 10, 20, and 50$\times 10^{-7}$  (A$^{2}$ cm$^{-2}$). 
%50,100,200,500,1000,2000,4000]/5000 
The contours in the right panels (b and d) indicate the current helicity density $h_z$ of  $\pm$5, 10, 20, 50, 100, and 200$\times10^{-4} $ (G$^{2}$ cm$^{-1}$). The  red (green) contours refer to the positive (negative) sign.    
The gray scale shows the corresponding magnetic free energy density  $\rho_{fh}$ (left) and  difference  quantity $\frac{1}{4\pi}{\bf B}_{nh}\cdot{\bf B}_{ph}$ (right) shown in Figure \ref{Fig:maen11158-14-15}.
}
\label{Fig:maen11158-curhel}
\end{center}
\end{figure*}
}

\section{Conclusion and Discussions}

We have presented the magnetic energy density parameters inferred from  photospheric vector magnetograms of the recurrent Active Region NOAA 6580-6619-6659 and 11158. This provided the opportunity of analyzing the storage and evolution of magnetic energy in active regions. After this analysis, the main results are as follows.

1) Observations of photospheric vector magnetic fields provide important information on the distribution of the free magnetic energy density in the lower solar atmosphere. The free magnetic energy density is comprised of the two terms $\frac{1}{8\pi}{B}_{nh}^2$ and $\frac{1}{4\pi}{\bf B}_{nh}\cdot{\bf B}_{ph}$. The first term refers to non-potential magnetic fields and the second to the relationship between the non-potential and potential fields. This  means that the change in the photospheric free magnetic energy density not only depends on  the non-potential component of magnetic field but also on the relationship with the potential field.

\begin{figure*}[htbp]
\begin{center}
\includegraphics[width=0.5\textwidth,clip=2]{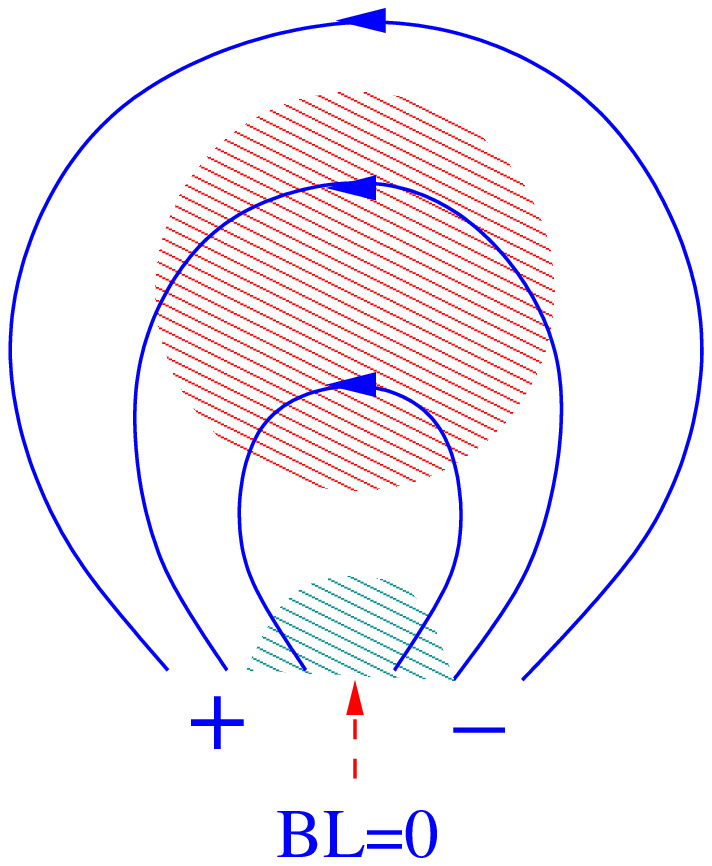}
\caption{A simplified schematic of $\frac{1}{4\pi}{\bf B}_{nh}\cdot{\bf B}_{ph}$ with height above the photospheric magnetic neutral line in the highly sheared active region as viewed from the side. The red (blue) shade shows the positive (negative) sign area. The red arrow indicates the photospheric magnetic neutral line in the active region.}
\label{Fig:shearv}
\end{center}
\end{figure*}

{ 2) The term $\frac{1}{4\pi}{\bf B}_{nh}\cdot{\bf B}_{ph}$ is an important quantity for understanding the degree of magnetic shear in the active regions.} The negative sign of $\frac{1}{4\pi}{\bf B}_{nh}\cdot{\bf B}_{ph}$  tends to occur in areas of highly sheared magnetic fields in the active regions.
Moreover, in the strongly sheared areas $(\frac{1}{8\pi}{B}_{nh}^2+\frac{1}{4\pi}{\bf B}_{nh}\cdot{\bf B}_{ph})<0$ relative to the highly sheared magnetic fields defined as a negative energy region can form in  delta active regions.  If the inclination angles between potential and non-potential magnetic lines of force decrease with height in the solar atmosphere of the active regions, the term $\frac{1}{4\pi}{\bf B}_{nh}\cdot{\bf B}_{ph}$ will change sign from negative to positive with the increase of height near the magnetic neutral lines in the active { regions.  This is consistent with the idea that some
%, which implies an observational evidence that the amount of 
free magnetic energy is stored} in the high solar atmosphere of active regions where powerful flares might be triggered. The trigger of powerful flares and CMEs in active regions, such as NOAA 6659 and 11158, was analyzed by some authors \citep[cf.][]{Schmieder94, Zhang94b,Yan95a,Wang97,Choudhary00,  Liu12,Schrijver11,Sun12,Chintzoglou13, Jiang13,Shen13, Sorriso-Valvo15}. A very simplifed picture on the relationship $\frac{1}{4\pi}{\bf B}_{nh}\cdot{\bf B}_{ph}$ with height in active regions is shown  in Figure \ref{Fig:shearv}. 

3) The mean photospheric magnetic energy in the active regions obviously changes before some powerful solar flares. { This might reflect the release of stored free magnetic energy powering flares, even if the free energy density can be negative}  in the photosphere in localized areas of some active regions.

From the analysis of the magnetic energy density of active regions in the lower solar atmosphere, we would like to discuss the following questions.

The analysis of magnetic shear in active regions is normally based on the inclination angle between the observed and the potential transverse magnetic field. The magnetic shear provides some information about the triggering of solar flares and CMEs, although it cannot provide all information about the free magnetic energy in  flare- or CME-producing regions even in the lower solar atmosphere because the real configuration of the magnetic fields in active regions is more complex.

The transverse and longitudinal components of the magnetic fields are inferred from the Stokes parameters Q, U, and V, respectively, with different calibration coefficients. The accuracy of the measurements of different components of the vector magnetic fields is still a basic problem for determining the photospheric  free magnetic energy density in solar active regions.

It should be noted that the  transverse potential field is a fictitious quantity inferred from the observational longitudinal field, so that its strength is a reference value in calculating the magnetic energy density of active regions in the photosphere.
The transverse components of potential magnetic fields can be calculated by the extrapolating the longitudinal components of the observed magnetic fields.  The estimation of the non-potential components of the fields also obviously depends on the measured  longitudinal fields and on the choice of extrapolation methods for the horizontal components of the potential fields.

\acknowledgments
The author would like to thank the anonymous referee for the suggestions and comments that helped improve the manuscript.
This study is supported by grants from the National Natural Science Foundation (NNSF) of China under the project grants 10921303, 11221063, 11673033 and 41174153.

\end{article} 

\end{document}